\documentclass[preprint,showpacs,preprintnumbers,amsmath,amssymb]{revtex4}%
\usepackage{graphicx}
\usepackage{dcolumn}
\usepackage{bm}
\usepackage{amsmath}
\usepackage{epstopdf}
\usepackage{amsfonts}
\usepackage{amssymb}%
\providecommand{\U}[1]{\protect\rule{.1in}{.1in}}
\newcommand{\be}{\begin{equation}}
\newcommand{\en}{\end{equation}}
\newcommand{\bea}{\begin{eqnarray}}
\newcommand{\ena}{\end{eqnarray}}
\begin{document}
\title{Hamilton-Jacobi approach for quasi-exponential inflation: predictions and constraints after Planck 2015 results}
\author{Nelson Videla}
\email{nelson.videla@ing.uchile.cl}
\affiliation{Departamento de F\'{\i}sica, FCFM, Universidad de Chile, Blanco Encalada 2008, Santiago, Chile}
\date{\today}

\begin{abstract}
In the present work we study the consequences of considering an inflationary universe
model in which  the Hubble rate has a
quasi-exponential dependence in the inflaton field, given by $H(\phi)=H_{inf}\exp \left[\frac{\frac{\phi}{m_p}}{p\left(1+\frac{\phi}{m_p}\right)}\right]$. We analyze the
inflation dynamics under the Hamilton-Jacobi approach, which allows us to consider $H(\phi)$, rather than $V(\phi)$, as the fundamental quantity to be specified. By comparing the
theoretical predictions of the model together with the allowed
contour plots in the $n_s-r$ plane and the amplitude of primordial scalar perturbations from the latest Planck data,
the parameters charactering this model are constrained. The model predicts values for the tensor-to-scalar ratio $r$ and for the running of the scalar spectral index $dn_s/ d\ln k$ consistent with the current bounds imposed by Planck, and we conclude
that the model is viable.
\end{abstract}

\maketitle



\section{Introduction}

Inflation has become the most acceptable paradigm that describes the physics of the very early universe. Besides of solving most of the shortcomings of the hot big-bang scenario, like the horizon, the flatness, and the
monopole problems \cite{R1,R106,R103,R104,R105,Linde:1983gd}, inflation also generates a causal mechanism to explain the large-scale structure (LSS) of the universe  \cite{R2,R202,R203,R204,R205}
and the origin of the anisotropies observed in the cosmic microwave background (CMB) radiation \cite{astro,astro2,astro202,Hinshaw:2012aka,Ade:2013zuv,Ade:2013uln,Ade:2015xua,Ade:2015lrj}, since primordial density perturbations may be sourced from quantum fluctuations of the inflaton scalar field during the inflationary expansion.

Several representative inflationary models have been studied within the framework of the so-called slow-roll approximation \cite{Lyth:2009zz},
where the kinetic term of the inflaton field
is much smaller than the potential energy, i.e. $\dot{\phi}^2\ll V(\phi)$, together with the approximation $\left|\ddot{\phi}\right|\ll H \left|\dot{\phi}\right|$. Moreover, in this approach the full shape of the inflaton potential is considered in order to identify the value of the scalar field at the end of inflation and hence the value of the scalar field when the largest scales observable today cross the Hubble radius. Upon comparison to the current cosmological and astronomical observations, specially those related with the CMB
temperature anisotropies, it is possible to constrain several inflation models. Particularly, the constraints in the $n_s-r$ plane
give us the predictions of a number of representative inflationary
potentials. Recently, the Planck
collaboration has published new data of enhanced precision of
the CMB anisotropies \cite{Ade:2015lrj}. Here, the Planck full mission
data has improved the upper bound on the tensor-to-scalar ratio
$r_{0.002} < 0.11$($95\%$ CL) which is similar to obtained from  \cite{Ade:2013uln}, in which
$r < 0.12$ ($95\%$ CL). From the particle physics point of view, it is natural to
begin by specifying the functional form of the potential.
However, even for simple choices, such as exponential \cite{Lucchin:1984yf}, constant \cite{R1}
or power-law potentials \cite{Linde:1983gd}, it is not possible to go further analytically. An alternative way is
to specify the time-dependence of the scale factor $a(t)$. Following Refs.\cite{Barrow:1993zq, Rendall:2005if, Barrow:2006dh}. exact solutions can also be found in the scenario of intermediate inflation. In this inflationary model the scale factor evolves as $a(t)\sim \exp\left(At^f\right) $, where $A$ and $f$ are two constant parameters such that $A>0$ and $0<f<1$. The expansion rate of this scale factor is slower than de Sitter inflation \cite{R1}, for which $a(t)\sim \exp(H t)$, where $H$ is the Hubble rate, which is a contant, but faster than power-law inflation, $a(t)\sim t^n$ \cite{Lucchin:1984yf} , where $n>1$.

Alternative to the slow-roll approximation, there is another method for studying inflation known as the
Hamilton-Jacobi approach \cite{Salopek:1990jq, Kinney:1997ne}. This formulation is a powerful way of rewriting the equations of motion for single-field inflation. It can be derived by considering the scalar field itself to be the time variable, which is possible during any epoch in which the scalar field evolves monotonically with time. It allows us to consider the Hubble rate or Hubble function $H(\phi)$ (from now on, not confuse with with the Hamiltonian function $H$), rather than the inflaton scalar potential $V(\phi)$, as the fundamental quantity to be specified. Because $H(\phi)$, unlike $V(\phi)$, is a geometric quantity, inflation is described more naturally in that language. The advantage of such an approach is that the form of the potential is readily deduced. As it was suggested in Refs.\cite{Lidsey:1991zp, Lidsey:1991dz, Hawkins:2000dq}, $H(\phi)$ should be viewed as the solution generating
function when analysing inflationary cosmologies. For instance, $H(\phi) \sim \exp(\phi)$ gives the power-law inflation model \cite{Lucchin:1984yf}. Furthermore, this formalism has been considered by Planck collaboration in order to reconstruct the inflaton potential  beyond slow-roll approximation \cite{Ade:2013uln, Ade:2015lrj}. For a representative list
of recent inflation models studied under Hamilton-Jacobi formalism where several expressions for $H(\phi)$ have been considered, see Refs.\cite{delCampo:2012qb,Pal:2011dt,Aghamohammadi:2014aca,Villanueva:2015ypa,Villanueva:2015xpa,Sheikhahmadi:2016wyz}.

Following Ref.\cite {Pal:2011dt}, a phenomenological quasi-exponential Hubble rate $H(\phi)$ yielding an inflationary
solution was proposed to be
\begin{equation}
H(\phi)=H_{inf}\exp \left[\frac{\frac{\phi}{m_p}}{p\left(1+\frac{\phi}{m_p}\right)}\right],\label{Hexp}
\end{equation}
where $p$ is a dimensionless parameter, $m_p$ denotes the Planck mass, and $H_{inf}$ is a parameter
with dimensions of Planck mass. It is interesting to mention that this model presents an improvement in comparison to power-law inflation model \cite{Lucchin:1984yf}, because the first one addresses the graceful-exit problem of inflation and the value predicted for the tensor-to-scalar ratio was compatible with Seven-Year WMAP \cite{astro202}, being
supported by the current data available at that time.

The main goal of the present work is to study the realization of inflation by reconsidering the expression for the Hubble rate given
by Eq.(\ref{Hexp}), in the light of the recent Planck
results. We stress that our work is different to previous work \cite {Pal:2011dt} in three ways. Firstly, in this work we restrict ourselves only to the inflationary predictions of this models. Secondly, in the previous paper the authors did not used the contour plots in the $n_s-r$ and $n_s-d n_s/d\ln k$ planes to constrain the parameters of the model they studied. Finally, in our work here we make use the latest data from Planck, not available at that time, to put bounds on the parameters of the model. We will show that our results are modified
compared to \cite{Pal:2011dt} using the Planck results. By comparing the
theoretical predictions of the model together with the allowed
contour plots in the $n_s-r$ plane, the model predicts a value for the tensor-to-scalar ratio $r$ detectable
by Planck, and we conclude that the model is viable.

We organize our work as follows: After this introduction, in the next section
we summarize the dynamics of
inflation in the Hamilton-Jacobi formalism. In the third section we analyze the inflation dynamics of the Hubble rate given by Eq.(\ref{Hexp}) in the
Hamilton-Jacobi framework, obtaining expressions for the scalar power spectrum, scalar spectral index, and tensor-to-scalar ratio
in terms of the free parameters characterizing the model which are constrained by considering the Planck 2015 results, through the allowed contour plots in the $r-n_s$ plane and the amplitude of the scalar power spectrum. In section \ref{wi} we discuss a little further how in the warm inflation scenario the radiation-dominated phase is achieved without introducing the reheating phase for this quasi-exponential Hubble function. In the last section we finish with
our conclusions. We choose units so that $c=\hbar=1$.

\section{Hamilton-Jacobi approach to inflation}\label{branerew}

\subsection{Dynamics of inflation}

In the simplest model of inflation in Einstein's General Relativity is a classical homogeneous scalar field $\phi=\phi(t)$, named the inflaton field, which is introduced into the action. The properties of the scalar potential determine how inflation evolves. For a flat Friedman-Lema\^{i}tre-Robertson-Walker (FLRW) metric,
the Friedmann and acceleration equations become
\begin{equation}
H^2 =\frac{8 \pi}{3 m^2_p}\left(\frac{\dot{\phi}^2}{2}+V(\phi)\right),\label{H2}
\end{equation}
and
\begin{equation}
\frac{\ddot{a}}{a} =-\frac{4 \pi}{3 m^2_p}\left(\dot{\phi}^2-V(\phi)\right),\label{aa}
\end{equation}

respectively, where $m_p=1/G$ corresponds to the Planck mass.

Besides the Einstein equations, the field satisfies the Klein-Gordon equation in this FLRW universe
\begin{equation}
\ddot{\phi}+3H\dot{\phi}+V^{\prime}=0,\label{KG}
\end{equation}
where prime indicates derivative with respect to $\phi$, and dot a derivative with respect to cosmic time.

The Friedmann and Klein-Gordon equations are the basis to construct the Hamilton-Jacobi formulation. By combining Eqs.(\ref{H2}) and (\ref{KG}), we obtain the following expression
\begin{equation}
\dot{\phi}=-\left(\frac{m^2_p}{4 \pi}\right) H^{\prime}(\phi),\label{dotphi}
\end{equation}

which gives the relation between $\phi$ and cosmic time $t$. This allows us to write the Friedmann
equation in a first-order form, from which the inflaton potential $V(\phi)$ becomes
\begin{equation}
V(\phi)=\left(\frac{3m^2_p}{8\pi}\right)\left[H(\phi)^2-\frac{m^2_p}{4\pi}\left[H^{\prime}(\phi)\right]^2\right].\label{HJ}
\end{equation}

This last equation is the Hamilton-Jacobi equation \cite{Lyth:2009zz}. It allows us to consider $H(\phi)$, rather than $V(\phi)$, as the fundamental quantity to be specified. On the other hand, a relatiom $\frac{d a}{d \phi}=a \frac{H}{\dot{\phi}}$ with Eq.(\ref{dotphi})
yields a differential equation for $a(\phi)$, whose integration becomes
\begin{equation}
a(\phi)=\exp\left[-\frac{4\pi}{m^2_p}\int \,\frac{H^{\prime}(\phi)}{H(\phi)}\,d\phi\right]. \label{aphi}
\end{equation}

This equation implies that, once the functional form of a geometrical quantity $H(\phi)$ has been specified, the cosmological dynamics is determined. The advantage of such an approach is that the form of the
potential is readily deduced from Eq.(\ref{HJ}).

We can use the Hamilton-Jacobi formalism to write down a slightly different version of the slow-roll approximation, defining the Hubble hierarchy parameters $\epsilon_H$ and $\eta_H$ as \cite{Lyth:2009zz}
\begin{eqnarray}
\epsilon_H &\equiv & -\frac{d \ln H}{d \ln a} =\left(\frac{m^2_p}{4\pi}\right)\left(\frac{H^{\prime}(\phi)}{H(\phi)}\right)^2,\label{epsilon}\\
\eta_H & \equiv & -\frac{d \ln H^{\prime}}{d \ln a}= \frac{m^2_p}{4\pi}\frac{H^{\prime \prime}(\phi)}{H(\phi)}.\label{eta}
\end{eqnarray}

In the slow-roll limit, $\epsilon_H\,\,\rightarrow\,\,\epsilon$ and $\eta_H \,\,\rightarrow\,\, \eta -\epsilon$, where $\epsilon$ and $\eta$ are
the usual slow-roll parameters.  By using Eq.(\ref{epsilon}), the acceleration equation (\ref{aa}) is rewritten as
\begin{equation}
\frac{\ddot{a}}{a}=H^2\left(1-\epsilon_H\right).\label{newacc}
\end{equation}
During inflation $\epsilon_H$
satisfies the condition $\epsilon_H<1$, and the inflationary expansion ends when $\epsilon_H$ becomes one.

On the other hand, the number of $e$-folds between the Hubble-radius crossing and the end of inflation yields
\begin{equation}
N(\phi) \equiv \int_{t_{*}}^{t_{end}}\,H\,dt=\left(\frac{4\pi}{m^2_p}\right)\int_{\phi_{end}}^{\phi_{*}}\,\frac{H(\phi)}{H^{\prime}(\phi)}\,d\phi=\int_{\phi_{end}}^{\phi_{*}}\,\frac{1}{\epsilon_H}\frac{H^{\prime}(\phi)}{H(\phi)}\,d\phi\label{Nfolds}
\end{equation}
where $\phi_{*}$ and $\phi_{end}$ are the values of the scalar field when the cosmological scales cross the Hubble-radius and at the end of inflation, respectively. The last value is found by $\epsilon_H(\phi_{end})=1$.

\subsection{Attractor behavior}

The Hamilton-Jacobi approach is usefull to show that all possible inflationary trajectories will rapidly converge to a common  attractor solution, if they are sufficiently close to each other initially. This is exactly the behaviour that one expect within the slow-roll approximation, but the proof do not use of that approximation. Suppose that $H_0(\phi)$ is any solution to Eq.(\ref{HJ}), inflationary or not. If we add to this solution a linear homogeneous perturbation $\delta H(\phi)$, the attractor behaviour will be satisfied if $\frac{\delta H(\phi)}{H_0(\phi)}$ tends quickly to zero as $\phi$ evolves \cite{Liddle:1994dx}. Replacing $H(\phi)=H_0(\phi)+\delta H(\phi)$ in Eq.(\ref{HJ}) and linearizing, we have that
\begin{equation}
\delta H(\phi)\simeq \frac{1}{3}\left(\frac{m^2_p}{4\pi}\right)\frac{H^{\prime}_0(\phi)}{H_0(\phi)}\,\delta H^{\prime}_0(\phi).
\end{equation}
Integrating last expression we get
\begin{equation}
\delta H(\phi)=\delta H(\phi_i) \exp\left(\int_{\phi}^{\phi_i}\,\frac{3}{\epsilon_H}\frac{H^{\prime}_0(\phi)}{H_0(\phi)}\right)\,d\phi, \label{attractor}
\end{equation}
where $\delta H(\phi_i)$ is the initial value of the perturbation at $\phi=\phi_i$. Knowing $H(\phi)$, it is possible to study the behaviour
of perturbation $\delta H(\phi)$.

In the next section we will give a review of cosmological perturbations and use Hubble hierarchy parameters for describing scalar and tensor perturbations.

\subsection{Cosmological perturbations}

We consider the gauge invariant quantity $\zeta=-\psi-H\frac{\delta \rho}{\dot{\rho}}$. Here, $\zeta$ is defined on slices of uniform density and reduces to the curvature perturbation $\mathcal{R}$ at super-horizon scales. A fundamental
feature of $\zeta$ is that it is nearly constant on super-horizon scales \cite{Riotto:2002yw}, and in fact this property does not depend on the gravitational field equations \cite{Wands:2000dp}. Therefore, at super-horizon scales we have that $\mathcal{R}=H\frac{\delta \phi}{\dot{\phi}}$, where $\left|\delta \phi\right|=H/2\pi$. In this way, the power spectrum
of scalar perturbations is given by \cite{Lyth:2009zz, Bassett:2005xm}
\begin{equation}
\mathcal{P}_{\mathcal{R}}(k)=\frac{H^2}{\dot{\phi}^2}\left(\frac{H}{2\pi}\right)^2_{k=aH}.\label{AS}
\end{equation}
This perturbation is evaluated at Hubble radius crossing $k = aH$ during inflation.

Important observational quantities are not only the amplitude of the primordial curvature perturbations but also the scalar spectral index which represents the scale dependence of the power spectrum, defined by
\begin{equation}
n_s-1\equiv \frac{d \ln \mathcal{P}_{\mathcal{R}} }{d \ln k}.
\end{equation}
Thus, the scalar spectral index of the power spectrum (\ref{AS}) is given by
\begin{equation}
n_s-1=2\eta_H-4\epsilon_H, \label{ns}
\end{equation}
where $\epsilon_H$ and $\eta_H$ are the Hubble hierarchy parameters, given by Eqs.(\ref{epsilon}) and (\ref{eta}), respectively.

We also introduce the running of the scalar spectral index, which represents the scale dependence of the spectral index, by
$n_{run}=\frac{d n_s}{d \ln k}$, yielding
\begin{equation}
n_{run}=10 \epsilon_H \eta_H-8 \epsilon^2_H-2 \xi^2_H,\label{nrun}
\end{equation}
where $\xi^2_H$ is a third Hubble hierarchy parameter, defined by \cite{Lidsey:1995np}
\begin{equation}
\xi^2_H\equiv \frac{m^2_p}{4 \pi}\left(\frac{H^{\prime \prime \prime}(\phi)H^{\prime}(\phi)}{H(\phi)^2}\right).\label{xi2}
\end{equation}

On the other hand, the power spectrum of tensor perturbations generated from inflation is given by \cite{Lyth:2009zz, Bassett:2005xm}
\begin{equation}
\mathcal{P}_{\mathcal{T}}=\frac{16\pi}{m^2_p}\left(\frac{H}{2\pi}\right)^2_{k=aH}.\label{TS}
\end{equation}

As the cosmological parameter related to the primordial tensor perturbation, the ratio between the amplitude of the primordial tensor perturbation and that of the primordial curvature perturbation, the so-called tensor-to-scalar ratio, defined by $r\equiv \frac{\mathcal{P}_{\mathcal{R}}}{\mathcal{P}_{\mathcal{T}}}$, becomes
\begin{equation}
r=4\epsilon_H.\label{rH}
\end{equation}

Additionally, by combining Eqs.(\ref{dotphi}) and (\ref{rH}), we obtain the Lyth bound \cite{Lyth:1996im}, which relates the tensor-to-scalar ratio and
the evolution of the scalar inflaton field
\begin{equation}
\frac{\Delta \phi}{m_p}=\frac{1}{4\sqrt{\pi}}\int_{0}^{N}\,\sqrt{r}\,dN.\label{lyth}
\end{equation}

This means that the tensor-to-scalar ratio measures (up to order-one constants) the distance that the inflaton
field $\phi$ traveled in field space during inflation. For detectable $r$, this implies $\Delta \phi \sim m_p$.

Up to now, the basis of the Hamilton-Jacobi formalism has been presented. In next section, in order to get a specific result, we are going to introduce the quasi-exponential form for Hubble rate given by Eq.(\ref{Hexp}).

\section{Hamilton-Jacobi approach for quasi-exponential inflation}\label{natbra}

\subsection{Dynamics of inflation}

In this section we describe an inflationary model by using the quasi-exponential generating function $H(\phi)$ given by Eq.(\ref{Hexp}). By combining Eqs.(\ref{Hexp}) and (\ref{dotphi}) we get that
\begin{eqnarray}
&\exp\left[-\frac{\frac{\phi}{m_p}}{p\left(1+\frac{\phi}{m_p}\right)}\right] p\left(1+\frac{\phi}{m_p}\right)\left[1+p+2p^2+p(1+4p)\frac{\phi}{m_p}+2p^2\left(\frac{\phi}{m_p}\right)^2\right] \nonumber\\
&-\exp\left[-\frac{1}{p}\right] \textup{Ei}\left[x(\phi)\right]=-\frac{3p^2}{2\pi}H_{inf}t,\label{phisol}
\end{eqnarray}
where $\textup{Ei}\left[x(\phi)\right]$ denotes the Exponential Integral function \cite{arfken}, given by the integral
\begin{equation}
\textup{Ei}(x)=-\int_{-x}^{\infty}\,\frac{\exp(-z)}{z}\,dz,
\end{equation}
with $x(\phi)=\frac{1}{p\left(1+\frac{\phi}{m_p}\right)}$. This latter expression yields the inflaton field as function of cosmic time. Additionally, the scale factor $a(t)$ turns out to be
\begin{equation}
a(\phi)=a_i \exp\left[-4\pi p\left(\left[\frac{\phi}{m_p}+\left(\frac{\phi}{m_p}\right)^2+\left(\frac{\phi}{m_p}\right)^3\right]-\left[\frac{\phi_i}{m_p}+\left(\frac{\phi_i}{m_p}\right)^2+\left(\frac{\phi_i}{m_p}\right)^3\right]\right)\right],\label{asolphi}
\end{equation}
where $a_i$ denotes the value of the scale factor when the inflaton field has the value $\phi_i$, i.e., $a_i=a(\phi_i)$. In order to have
an inflationary solution, the condition $\phi<\phi_i$ must be satisfied, which means that the inflaton starts to rolling down the
potential at large values of $\phi_i$.

The form of the potential is readily deduced from Eqs.(\ref{Hexp}) and (\ref{HJ}), which results to be given by
\begin{eqnarray}
&V(\phi)=V_0\frac{\exp\left[\frac{\frac{2\phi}{m_p}}{p\left(1+\frac{\phi}{m_p}\right)}\right] }{\left(1+\frac{\phi}{m_p}\right)^4}\bigg[\left(4 \pi p^2-1\right)+16\pi p^2\frac{\phi}{m_p}+24\pi p^2\left(\frac{\phi}{m_p}\right)^2+16\pi p^2\left(\frac{\phi}{m_p}\right)^3\nonumber\\
&+4\pi p^2\left(\frac{\phi}{m_p}\right)^4\bigg],\label{VVphi}
\end{eqnarray}
where $V_0=\frac{3H_{inf}^2m^2_p}{32 \pi^2 p^2}$. For sake of comparison, in the slow-roll approximation, $\dot{\phi}^2\ll V(\phi)$ and $\left|\ddot{\phi}\right|\ll H \left|\dot{\phi}\right|$, the inflaton potential becomes
\begin{equation}
V(\phi)\simeq \frac{3 H^2_{inf}m^2_p}{8\pi} \exp\left[\frac{\frac{2\phi}{m_p}}{p\left(1+\frac{\phi}{m_p}\right)}\right].
\end{equation}

For this model the Hubble hierarchy parameters $\epsilon_H$ and $\eta_H$ become
\begin{equation}
\epsilon_H(\phi)=\frac{1}{4\pi p^2\left(1+\frac{\phi}{m_p}\right)^4},\label{eh}
\end{equation}
and
\begin{equation}
\eta_H(\phi)=-\frac{\left(-1+2p+2p\frac{\phi}{m_p}\right)}{4\pi p^2\left(1+\frac{\phi}{m_p}\right)^4},\label{etah}
\end{equation}
respectively.

From the condition $\epsilon_H(\phi_{end})=1$, we obtain the value of the inflaton field at the end of the inflationary expansion, yielding
\begin{equation}
\phi_{end}=\left(\frac{1}{\sqrt{2p}\,\pi^{1/4}}-1\right)m_p.\label{phiend}
\end{equation}
Restricting ourselves only to positive incursion of the inflaton field trough the potential, the allowed range for $p$ becomes $0 < p < \frac{1}{2\sqrt{\pi}}\approx 0\textup{.}282$.

The number of inflationary $e$-folds between the values of the scalar field when a given perturbation scale leaves the Hubble-radius and at the end of inflation, can be computed from Eqs.(\ref{Hexp}), (\ref{Nfolds}), and (\ref{phiend}), resulting in
\begin{equation}
N=\frac{4\pi p}{3}-\frac{\sqrt{2}\pi^{1/4}}{3\sqrt{p}}+4\pi p \frac{\phi_{*}}{m_p}+4\pi p \left(\frac{\phi_{*}}{m_p}\right)^2+\frac{4\pi p}{3}\left(\frac{\phi_{*}}{m_p}\right)^3.\label{efoldsV}
\end{equation}
By solving Eq.(\ref{efoldsV}) for $\phi_{*}$, we may obtain the value of the scalar field at the time of Hubble-radius crossing, giving
\begin{equation}
\phi_{*}=\left[\frac{\left(3N \pi^2p^2+\sqrt{2}\pi^{9/4}p^{3/2}\right)^{1/3}}{2^{2/3}\pi p}-1\right]m_p.\label{ycmb}
\end{equation}

As we shall see later on, the several inflationary observables will be evaluated at the value of the inflaton field given by Eq.(\ref{ycmb}).

\subsection{Attractor behavior}

\begin{figure}[th]
{\hspace{-2
cm}\includegraphics[width=3.3 in,angle=0,clip=true]{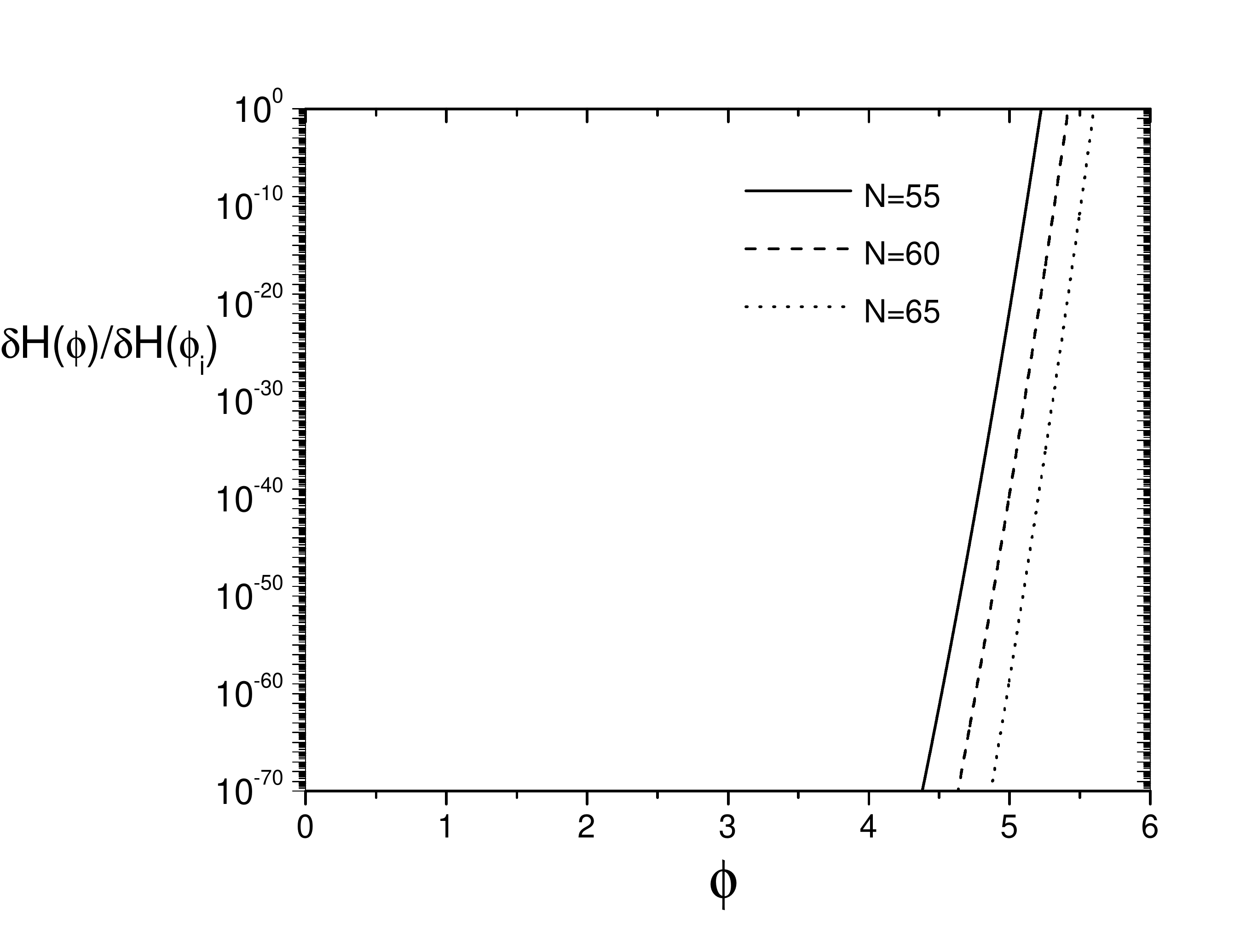}
}
{\vspace{-0.5 cm}\caption{Plot of the perturbation $\delta H(\phi)/H(\phi_i)$ as function of inflaton field $\phi$ for the
quasi-exponential Hubble rate. For this plot we
have used 3 different values for the number of $e$-folds $N$: the solid, dashed, and dotted lines correspond to $N=55,\,60$, and $65$, respectively. Additionally, we have used the values $p=0\textup{.}15$ and $\alpha=1\textup{.}5$.
 \label{delta}}}
\end{figure}

As final step of the analysis of background dynamic for this model, the attractor behavior of the solution is
considered. From Eqs.(\ref{Hexp}) and (\ref{attractor}), the solution for the perturbation $\delta H(\phi)$ yields
\begin{equation}
\frac{\delta H(\phi)}{\delta H(\phi_i)}=\exp\left(4\pi p \left[\left(1+\frac{\phi}{m_p}\right)^3-\left(1+\frac{\phi_i}{m_p}\right)^3\right]\right).\label{deltahs}
\end{equation}

In order to determine  the attractor behaviour quantitatively, we consider the initial value of the inflaton field to be $\phi_i=\alpha \phi_{*}$, with $\alpha >1$ and $\phi_{*}$ given by Eq.(\ref{ycmb}). Then, replacing $\phi_i$ into Eq.(\ref{deltahs}), the perturbation $\frac{\delta H(\phi)}{\delta H(\phi_i)}$
becomes
\begin{equation}
\frac{\delta H(\phi)}{\delta H(\phi_i)}=\exp\left(4\pi p \left[\left(1+\frac{\phi}{m_p}\right)^3-\left(1+\alpha \left[\frac{\left(3N \pi^2p^2+\sqrt{2}\pi^{9/4}p^{3/2}\right)^{1/3}}{2^{2/3}\pi p}-1\right]\right)^3\right]\right).\label{deltahN}
\end{equation}

Fig.\ref{delta} shows the plot of the perturbation $\delta H(\phi)/H(\phi_i)$ as function of inflaton field $\phi$. For this plot we
have used 3 different values for the number of $e$-folds $N$: the solid, dashed, and dotted lines correspond to $N=55,\,60$, and $65$, respectively. Additionally, we have used the values $p=0\textup{.}15$ and $\alpha=1\textup{.}5$.

For this quasi-exponential form of the Hubble function, the inflaton field decreases as time increases, therefore the exponential term
on the right-hand side of Eq.(\ref{deltahN}) decreases by passing time and tends to zero rapidly, then the
perturbation of the Hubble function vanishes, and the model has an attractive behavior.

\subsection{Cosmological perturbations}

Regarding the cosmological perturbations, the amplitude of the primordial curvature perturbation, using Eqs.(\ref{Hexp}) and (\ref{AS}), is found to be
\begin{equation}
\mathcal{P}_{\mathcal{R}}=\frac{4H^2_{inf}p^2}{m^2_p}\exp\left[\frac{2\frac{\phi}{m_p}}{p\left(1+\frac{\phi}{m_p}\right)}\right]\left(1+\frac{\phi}{m_p}\right)^4.\label{ASV}
\end{equation}

The scalar spectral index, using Eqs.(\ref{ns}), (\ref{eh}), and (\ref{etah}), becomes
\begin{equation}
n_s=1-\frac{\left(1+2p+2p\frac{\phi}{m_p}\right)}{2\pi p^2 \left(1+\frac{\phi}{m_p}\right)^4}.\label{nsV}
\end{equation}

Additionally, the running of the scalar spectral index $n_{run}$ is found to be
\begin{equation}
n_{run}=-\frac{\left(2+3p+3p\frac{\phi}{m_p}\right)}{4\pi^2 p^2 \left(1+\frac{\phi}{m_p}\right)^7}.\label{nsV}
\end{equation}

Finally, the tensor-to-scalar ratio can be obtained from Eqs.(\ref{rH}) and (\ref{eh}), yielding
\begin{equation}
r=\frac{1}{\pi p^2\left(1+\frac{\phi}{m_p}\right)^4}.\label{rrV}
\end{equation}

After evaluating these inflationary observables at the value of the scalar field when a given perturbation scale leaves the Hubble-radius, given by (\ref{ycmb}), we may compare the theoretical predictions of our model with the observational data in order to obtain constraints on the parameters that characterize it.

\begin{figure}[th]
{\hspace{-3
cm}\includegraphics[width=4.8 in,angle=0,clip=true]{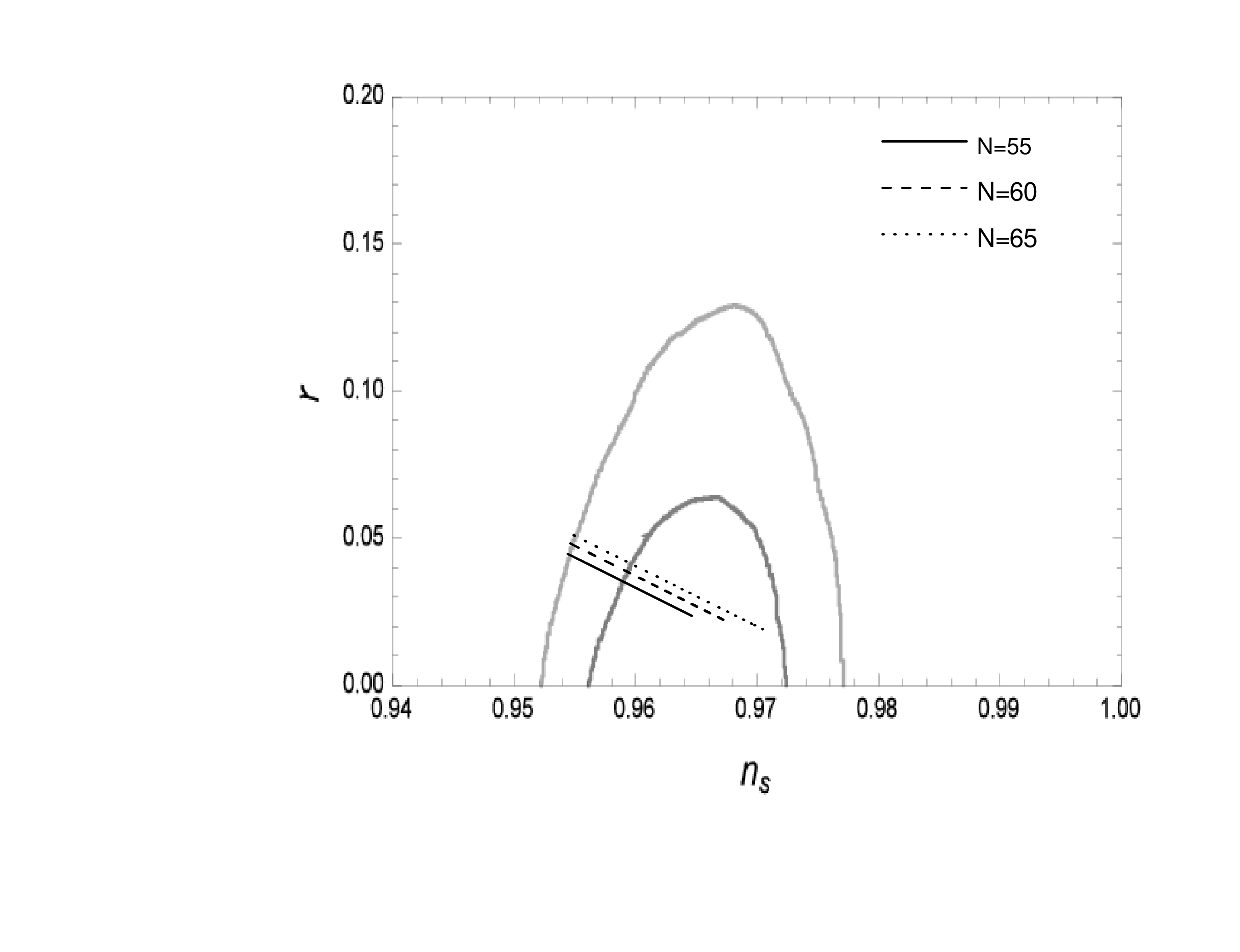}}
{\vspace{-1.5cm}\caption{Plot of the tensor-to-scalar ratio $r$ versus the scalar spectral index $n_s$ for the
quasi-exponential Hubble rate. Here, we have considered the
two-dimensional marginalized joint confidence contours for $(n_s
,r)$, at the $68\%$ and $95\%$ CL, from the latest Planck data \cite{Ade:2015lrj}. In this plot we
have used 3 different values for the number of $e$-folds $N$: the solid, dashed, and dotted lines correspond to $N=55,\,60$, and $65$, respectively.
 \label{rns}}}
\end{figure}

The amplitude of the primordial curvature perturbation, the scalar spectral index, the running of the scalar spectral index, and the tensor-to-scalar ratio, evaluated at the Hubble-radius crossing $k=aH$, become
\begin{eqnarray}
\label{AsN}
\mathcal{P}_{\mathcal{R}} &=& \frac{H^2_{inf}}{2^{2/3}\pi^{4/3}m^2_p}\exp\left[\frac{2}{p}-\frac{2^{5/3}\pi^{1/3}}{\sqrt{p}\left(3N\sqrt{p}+\sqrt{2}\pi^{1/4}\right)^{1/3}}\right]\left(3N\sqrt{p}+\sqrt{2}\pi^{1/4}\right)^{4/3},\\
\label{nsN}
n_s &=&  1-\frac{4\sqrt{p}}{3N\sqrt{p}+\sqrt{2}\pi^{1/4}}-\frac{2^{5/3}\pi^{1/3}}{\left(3N\sqrt{p}+\sqrt{2}\pi^{1/4}\right)^{4/3}} ,\\
\label{runn}
 n_{run} &=& -\frac{4p}{\left(3N\sqrt{p}+\sqrt{2}\pi^{1/4}\right)^2}\left[3+\frac{2^{5/3}\pi^{1/3}}{\sqrt{p}\left(3N\sqrt{p}+\sqrt{2}\pi^{1/4}\right)^{1/3}}\right],\\
 \label{rN}
 r &=& \frac{2^{8/3}\pi^{1/3}}{\left(3N\sqrt{p}+\sqrt{2}\pi^{1/4}\right)^{4/3}}.
\end{eqnarray}

The first constraint on the parameters of this model can easily found from Eq.(\ref{AsN}), because we may write the $H_{inf}$ parameter in terms of the amplitude of the scalar power spectrum, obtaining
\begin{equation}
\label{HN}
H_{inf}=\frac{2^{1/3}\pi^{2/3}p\sqrt{\mathcal{P}_{\mathcal{R}}}m_p}{\sqrt{p}\left(3N\sqrt{p}+\sqrt{2}\pi^{1/4}\right)^{2/3}}\exp\left[-\frac{1}{p}+\frac{2^{2/3}\pi^{1/3}}{\sqrt{p}\left(3N\sqrt{p}+\sqrt{2}\pi^{1/4}\right)^{1/3}}\right].
\end{equation}

The trajectories in the $n_s-r$ plane for the model studied here may be generated by plotting  Eqs.(\ref{nsN}) and (\ref{rN}) parametrically. In particular, we have obtained three different curves by fixing the number of $e$-folds to $N=55,\,60$, and $65$, and plotting with respect to the parameter $p$ in the range $0 < p < \frac{1}{2\sqrt{\pi}}$, obtained by considering a positive incursion of the inflaton field trough the potential, which gives an upper bound for $p$. The Fig.(\ref{rns}) shows the plot of the tensor-to-scalar ratio $r$ versus the scalar spectral index $n_s$ for the quasi-exponential Hubble rate. Here, we have considered the
two-dimensional marginalized joint confidence contours for $(n_s
,r)$, at the $68\%$ and $95\%$ CL, from the latest Planck data \cite{Ade:2015lrj}. We can determinate numerically from Eq.(\ref{rN})  that, by fixing $N$, the tensor-to-scalar ratio decreases as the parameter $p$ is increasing. In this way, the allowed contour plots in the $n_s-r$ plane impose a strong constraint on the lower bound for $p$. This lower bound for the dimensionless parameter $p$, for each $r(n_s)$ curve, may be inferred by finding the points when the trajectory enters the $95\%$ CL region from Planck.
The trajectory for $N=55$ enters to joint $95\%$ CL region in the $n_s$ - $r$ plane for $p>0\textup{.}104$. On the other hand,
for $N=60$, the trajectory enters to the $95\%$ CL region for $p>0\textup{.}078$ . Finally, for $N=65$ enters to  the joint $95\%$ CL
 for $p>0\textup{.}061$. On the other hand, by Eq.(\ref{HN}), the constraints on $p$ already obtained, and the observational value for amplitude of the scalar power spectrum $\mathcal{P}_{\mathcal{R}}\simeq 2 \times 10^{-9}$ \cite{Ade:2015lrj}, me may obtain the allowed range for $H_{inf}$ for each value of $N$. For $N=55$, this constraint becomes $3\text{.}883\times 10^{-9}\,m_p< H_{inf} < 4\text{.}866\times 10^{-7}\,m_p$, for $N=60$ we have that $2\text{.}279\times 10^{-10}\,m_p < H_{inf} < 4\text{.}472\times 10^{-7}\,m_p$,
 and finally, for $N=65$ the allowed range becomes $9\text{.}117\times 10^{-12}\,m_p < H_{inf} < 4\text{.}141\times 10^{-7}\,m_p$. Table (\ref{T1}) summarizes the constraints obtained on $p$ and $H_{inf}$ using the last data of Planck.

\begin{table}
\centering
\begin{tabular}{|c|c|c|}
\hline
$N$ & constraint on $p$ & constraint on $H_{inf}$ \\
\hline
55 & $0\textup{.}104<p<0\textup{.}282$ & $3\text{.}883\times 10^{-9}\,m_p< p < 4\text{.}866\times 10^{-7}\,m_p$ \\
\hline
60 & $0\textup{.}078<p<0\textup{.}282$ & $2\text{.}279\times 10^{-10}\,m_p < p < 4\text{.}472\times 10^{-7}\,m_p$ \\
\hline
65 & $0\textup{.}061<p<0\textup{.}282$ & $9\text{.}117\times 10^{-12}\,m_p < p < 4\text{.}141\times 10^{-7}\,m_p$ \\
\hline
\end{tabular}
\caption{Results for the constraints on the parameters $p$ and
$H_{inf}$ for the quasi-exponential form for the Hubble rate, using the last data of Planck.} \label{T1}
\end{table}

\begin{figure}[th]
{\hspace{-2
cm}\includegraphics[width=3.8 in,angle=0,clip=true]{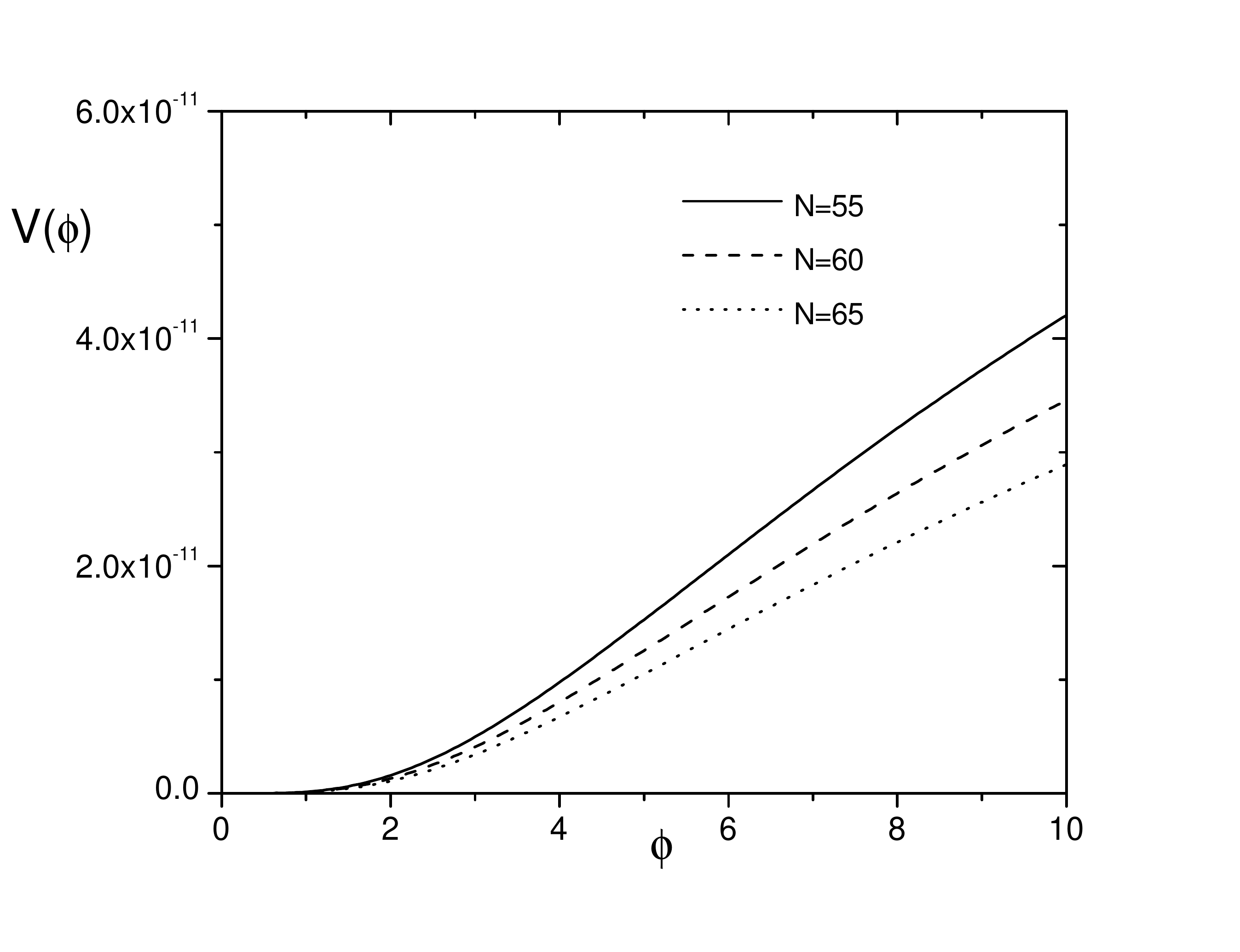}}
{\vspace{-1 cm}\caption{Plot of the scalar potential $V$ as function of the inflaton field. For this plot we
have used 3 different values for the number of $e$-folds $N$: the solid, dashed, and dotted lines correspond to $N=55,\,60$, and $65$, respectively. For all the three cases, we have used the values $p=0\textup{.}15$, which lies in the allowed range already obtained for ecah value of $N$, and $m_p=1$.
 \label{V}}}
\end{figure}

As we can see, using the latest Planck results, through the allowed contour plots in the $n_s-r$ plane and the amplitude of the scalar power spectrum, we were able to find the allowed range for $p$ and $H_{inf}$. Particularly, the allowed contour plots in the $n_s-r$ plane impose a strong constraint on the lower bound for $p$, which was not considered by the authors in the previous work \cite {Pal:2011dt}.

After replacing Eq.(\ref{HN}) into Eq.(\ref{VVphi}), we can plot the scalar potential $V$ as function of the inflaton field, as is shown in
Fig.(\ref{V}). We have plotted the inflaton potential for 3 different values for the number of $e$-folds $N$: the solid, dashed, and dotted lines correspond to $N=55,\,60$, and $65$, respectively, and fixing the value $p=0\textup{.}15$, which lies in the allowed range for each value of $N$ already obtained by using the Planck data. It is interesting to mention that this quasi-exponential form
of the Hubble rate presents a graceful-exit of inflation, however, the inflaton potential does not present a minimum, which raises the issue of how to address the problem of reheating in this model. A way to address this problem may be to study this model in the warm inflation scenario \cite{Berera:2008ar, BasteroGil:2009ec,Ramos:2016coz}, which has the attractiveness that it avoids the reheating period at the end of the accelerated expansion. In such as scenario, the dissipative effects are important, and radiation production
takes place at the same time as the expansion of the universe. When the universe heats up and becomes radiation dominated, inflation ends and the universe smoothly enters the radiation Big Bang phase. In section \ref{wi} we discuss a little further how in the warm inflation scenario the radiation-dominated phase is achieved without introducing the reheating phase for this quasi-exponential Hubble function.

\begin{figure}[th]
{\hspace{-2
cm}\includegraphics[width=3.3 in,angle=0,clip=true]{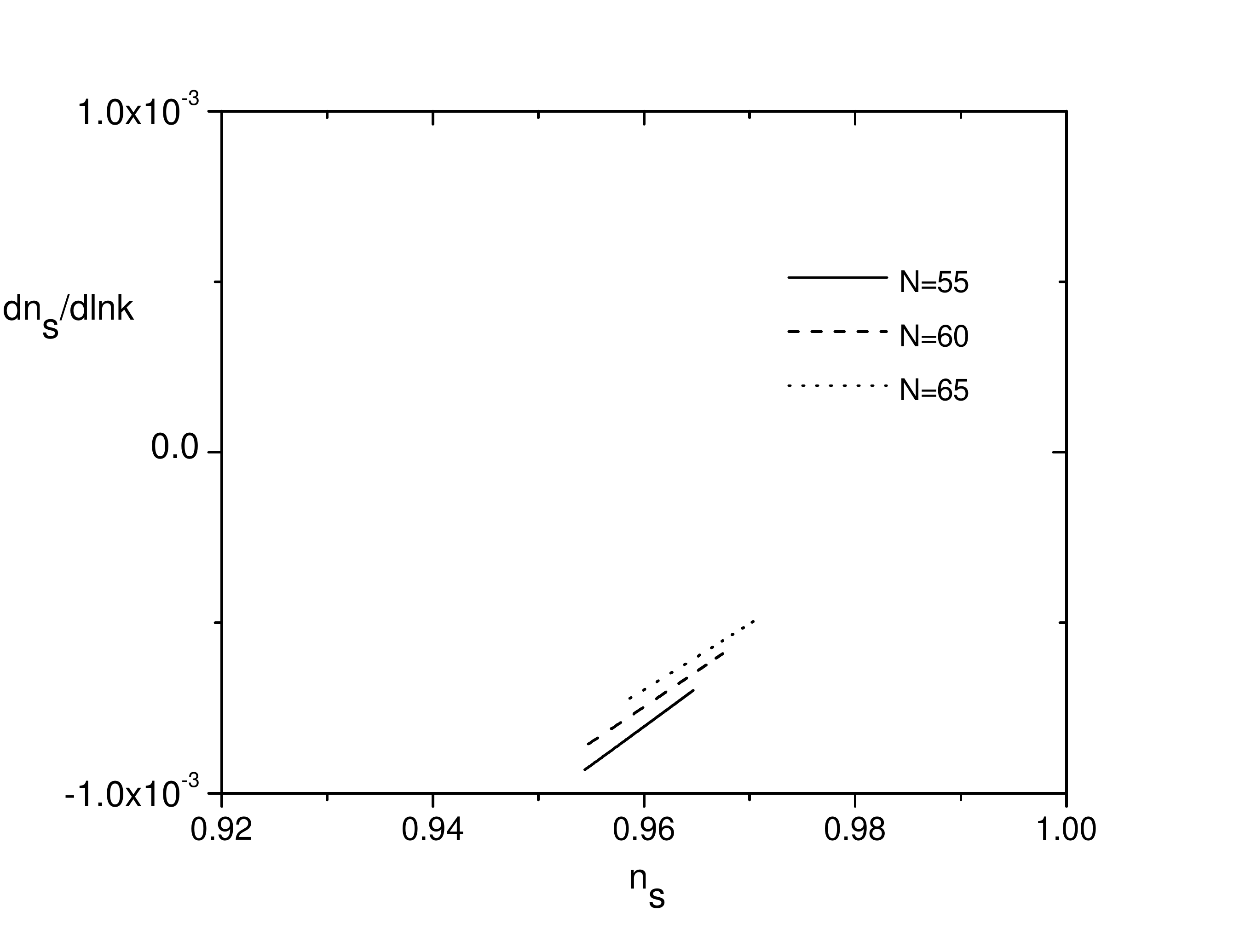}
\includegraphics[width=3.3 in,angle=0,clip=true]{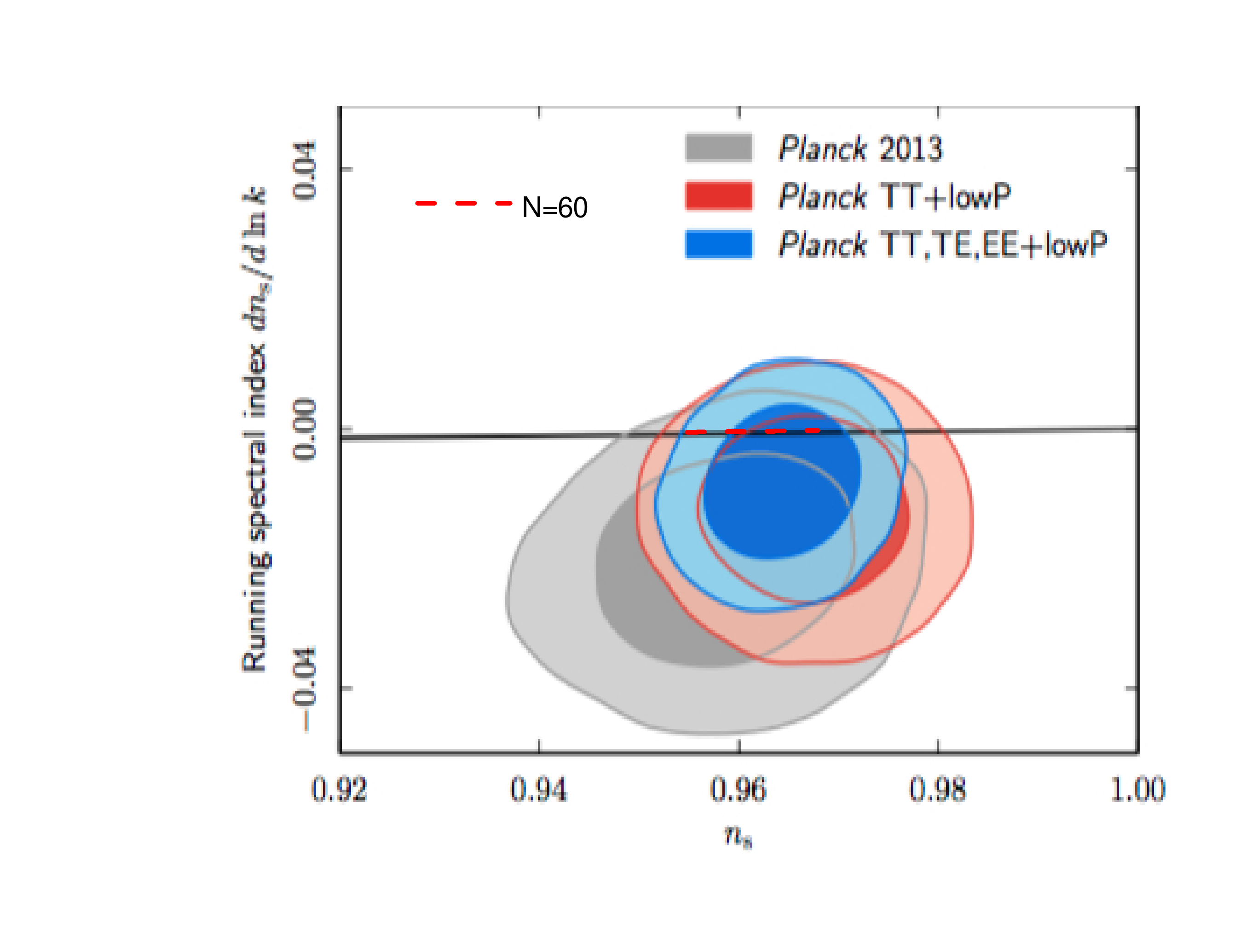}}
{\vspace{-1 cm}\caption{Left panel shows the plot of the running of the scalar spectral index $d n_s/ d\ln k$ versus the scalar spectral index $n_s$ for the
quasi-exponential Hubble rate. In this plot we
have used 3 different values for the number of $e$-folds $N$: the solid, dashed, and dotted lines correspond to $N=55,\,60$, and $65$, respectively. The right panel shows the two-dimensional marginalized joint confidence contours for $(n_s
,dn_s / d\ln k)$, at the $68\%$ and $95\%$ CL, in the presence of a non-zero tensor contribution, from the latest Planck data \cite{Ade:2015lrj} and the $\frac{d n_s}{d\ln k}(n_s)$ curve for $N=60$ (red-dashed line).
 \label{run}}}
\end{figure}

In order to determine the prediction of this model regarding the running of the spectral index, the trajectories in the $n_s-d n_s/d\ln k$ plane may be generated by plotting  Eqs.(\ref{runn}) and (\ref{rN}) parametrically. In particular, we have obtained three different curves by fixing the number of $e$-folds to $N=55,\,60$, and $65$, and plotting with respect to the parameter $p$ in the allowed range obtained for each value of $N$, which are shown in left panel of Fig.(\ref{run}). In order to compare the previous predictions with the observational data, the right panel of Fig.(\ref{run}), shows the
two-dimensional marginalized joint confidence contours for $(n_s,
d n_s/d\ln k)$, at the $68\%$ and $95\%$ CL, in the presence of a non-zero tensor contribution, from the latest Planck data \cite{Ade:2015lrj}. Given the indistinguishability of the curves in left panel, for the right panel we have only considered the curve corresponding to $N=60$ (red-dashed line) to compare with Planck data. The thin black line in right panel shows the prediction of single-field monomial inflation models with $50<N<60$. From both panels, we observe that all three $\frac{d n_s}{d\ln k}(n_s)$ curves lie  inside the $68\%$ as well as $95\%$ CL regions from Planck.

Finally, by replacing the expression
that we have found for the tensor-to-scalar ratio $r(N)$, expressed by Eq.(\ref{rN}), into Eq.(\ref{lyth}), the incursion of the inflaton field is found to be
\begin{equation}
\frac{\Delta \phi}{m_p}=\frac{1}{2^{2/3}\sqrt{p}\pi^{1/3}}\left[\left(3N\sqrt{p}+\sqrt{2}\pi^{1/4}\right)^{1/3}-\left(\sqrt{2}\pi^{1/4}\right)^{1/3}\right].
\end{equation}
In particular, by considering the constraint on $p$ for each value of $N$ already obtained, we get that
 $2\textup{.}623<\frac{\Delta \phi}{m_p}<3\textup{.}428$, $2\textup{.}727<\frac{\Delta \phi}{m_p}<3\textup{.}852$, and $2\textup{.}826<\frac{\Delta \phi}{m_p}<4\textup{.}267$, for $N=55$, $N=60$, and $N=70$, respectively. We note that the
 incursion of the inflaton field decreases as $p$ increases.

 \section{A first approach to warm inflation with a quasi-exponential Hubble function}\label{wi}

As we mentioned at previous section, the warm
inflation scenario, as opposed to the standard cold inflation, has the attractive feature that it
avoids the reheating period at the end of the accelerated expansion. During the evolution
of warm inflation dissipative effects are important, and radiation production takes place at
the same time as the expansion of the universe. The dissipative effects arise from a friction
term $\Gamma$ which accounts for the processes of the scalar field dissipating into a thermal bath. In addition,
in warm inflationary scenario the density perturbations arise
from thermal fluctuations of the inflaton and dominate over the quantum ones. In this form,
an essential condition for warm inflation to occur is the existence of a radiation component
with temperature $T>H$, since the thermal and quantum fluctuations are proportional to
$T$ and $H$, respectively \cite{Berera:2008ar, BasteroGil:2009ec,Ramos:2016coz}. When the universe heats up and becomes radiation dominated,
inflation ends and the universe smoothly enters in the radiation Big-Bang.

We start by considering a spatially flat FLRW universe
containing a self-interacting inflaton scalar field $\phi$ with energy density and pressure
given by $\rho_{\phi}=\dot{\phi}^2/2+V(\phi)$ and $P_{\phi}=\dot{\phi}^2/2-V(\phi)$, respectively,
and a radiation field with energy density $\rho_{\gamma}$. The corresponding Friedmann equations reads
\begin{equation}
H^2=\frac{8\pi}{3m^2_p}(\rho_{\phi}+\rho_{\gamma}), \label{Freq}
\end{equation}
and the dynamics of $\rho_{\phi}$ and $\rho_{\gamma}$ is described by the equations \cite{Berera:2008ar, BasteroGil:2009ec,Ramos:2016coz}
\begin{equation}
\dot{\rho_{\phi}}+3\,H\,(\rho_{\phi}+P_{\phi})=-\Gamma \dot{\phi}^{2},
\label{key_01}%
\end{equation}
and
\begin{equation}
\dot{\rho}_{\gamma}+4H\rho_{\gamma}=\Gamma \dot{\phi}^{2}, \label{key_02}%
\end{equation}
where the  dissipative coefficient $\Gamma>0$ produces the decay of the scalar
field into radiation. Recall that this
decay rate can be assumed  to be a function of the
temperature of the thermal bath $\Gamma(T)$, or a function of the
scalar field $\Gamma(\phi)$, or a function of $\Gamma(T,\phi)$ or
simply a constant\cite{Berera:2008ar, BasteroGil:2009ec,Ramos:2016coz}.

During warm inflation, the energy density related to
the scalar field predominates  over the energy density of the
radiation field, i.e.,
$\rho_\phi\gg\rho_\gamma$\cite{Berera:2008ar, BasteroGil:2009ec,Ramos:2016coz}, but even if small when compared to the inflaton energy density
it can be larger than the expansion rate with $\rho_{\gamma}^{1/4}>H$. Assuming thermalization, this translates roughly
into $T>H$, which is the condition for warm inflation to occur.

When $H$, $\phi$, and $\Gamma$ are slowly varying, which is a good
approximation during inflation, the production of radiation becomes quasi-stable, i.e., $\dot{\rho
}_{\gamma}\ll4H\rho_{\gamma}$ and $\dot{\rho}_{\gamma}\ll\Gamma\dot{\phi}^{2}%
$, see Refs.\cite{Berera:2008ar, BasteroGil:2009ec,Ramos:2016coz}. Then, the energy density of the radiation field becomes
\begin{equation}
4H\rho_{\gamma}\simeq \Gamma\,\dot{\phi}^{2}. \label{key_02n}%
\end{equation}
If we consider thermalization, then the energy density of the radiation field could be written as $\rho_{\gamma}=C_{\gamma}\,T^{4}$, where the constant  $C_{\gamma}=\pi^{2}\,g_{\ast}/30$. Here, $g_{\ast}$ represents the number
of relativistic degrees of freedom.
By combining Eqs.(\ref{Freq}) and (\ref{key_01}), the time derivative of the Hubble function is given by
\begin{equation}
\dot{H}(\phi)=-\left(\frac{4\pi}{m_p^2}\right)(1+Q)\dot{\phi}^2, \label{hdot}%
\end{equation}
where $Q$ is the dissipative ratio, defined as $Q\equiv \Gamma/3H$. In warm inflation, we can distinguish between two possible scenarios, namely the weak and strong dissipative regimes, defined as $Q\ll 1$ and $Q\gg 1$, respectively. In the weak dissipative regime, the Hubble damping is still the dominant term, however, in the strong dissipative regime, the dissipative coefficient $\Gamma$ controls the damped evolution of the inflaton field.

By expressing the time derivative of the Hubble function in terms of inflaton field derivative, the time derivative of
inflaton field becomes
\begin{equation}
\dot{\phi}=-\left(\frac{m_p^2}{4\pi}\right)\frac{H^{\prime}(\phi)}{(1+Q)}, \label{fdot}%
\end{equation}
which is the same expression found in \cite{Sayar:2017pam}, where warm inflation under the Hamilton-Jacobi formalism has been studied recently.

Introducing the dimensionless Hubble hierarchy parameter $\epsilon_H$, we write
\begin{equation}
\epsilon_H=\frac{1}{(1+Q)}\left(\frac{m_p^2}{4\pi}\right)\frac{H^{\prime\,2}(\phi)}{H^2(\phi)}.\label{ehw}%
\end{equation}

It is possible to find a relation between the energy densities $\rho_{\gamma}$ and $\rho_{\phi}$ by combining Eqs.(\ref{key_02n}), (\ref{fdot}), and (\ref{ehw}), so that
\begin{equation}
\rho_{\gamma}=\frac{Q}{2(1+Q)}\epsilon_H \rho_{\phi}.\label{rhos}%
\end{equation}

Warm inflation takes place when the parameter $\epsilon_H$ satisfies $\epsilon_H<1$. This condition given above
implies that during inflation the energy density of the inflaton field satisfies $\rho_{\phi}>\frac{2(1+Q)}{Q}\rho_{\gamma}$. Then,
at the end of inflation, when $\epsilon_H=1$, we have that $\rho_{\gamma}=\frac{Q}{2(1+Q)}\rho_{\phi}$. The universe stops inflating and heats up
to become radiation dominated at the time when $\rho_{\gamma}=\rho_{\phi}$. This is one of the most attractive features of warm inflation, since
provides a smooth transition to the radiation-dominated epoch without introducing a reheating epoch. Given that for the quasi-exponential Hubble function
the inflaton potential does not present a minimum, the dynamics for this model in warm inflation scenario provides a solution for
the problem of reheating. The perturbation dynamics of warm inflation with the quasi-exponential Hubble function deserves a more further analysis which goes beyond the scope of this work.

\section{Conclusions}\label{conclu}

To summarize, in this article we have studied an inflationary model the Hubble rate has an
quasi-exponential dependence in the inflaton field. We have studied the
inflation dynamics in the Hamilton-Jacobi formalism, in which the scalar field itself to be the time variable, which is possible during any epoch in which the scalar field evolves monotonically with time. It allows us to consider the Hubble function $H(\phi)$, rather than the inflaton potential $V(\phi)$, as the fundamental quantity to be specified. Because $H(\phi)$, unlike $V(\phi)$, is a geometric quantity, inflation is described more naturally in that language. This model is characterized by the dimensionless parameter $p$ and $H_{inf}$. In order to constraints our model, we have considered the amplitude of the primordial scalar perturbations, as well as the allowed contour plots in $n_s-r$ and $n_s-dn_s/ d\ln k$ planes from Planck 2015 data. First, in the $n_s-r$ plane we show the theoretical predictions of the model
for three different values of $e$-folds $N=55$, 60, and 65. By finding the points where each  $r(n_s)$ curve enters the joint
$95\%$ CL region, the allowed range for the $p$ parameter may be determined.  After that, using the constraint
for the amplitude of scalar perturbations we determined the allowed range for $H_{inf}$. In addition, in the $n_s-d n_s/ d\ln k$ plane we show that all the three $\frac{d n_s}{d\ln k}(n_s)$ curves for $N=55$, 60, and 65 lie  inside the $68\%$ as well as $95\%$ CL regions from Planck. This model predicts values for the tensor-to-scalar ratio $r$ and for the running of the scalar spectral index consistent with the current bounds imposed by Planck 2015, and we conclude
that the model is viable. As we mention before, this quasi-exponential form
of the Hubble rate presents a graceful-exit of inflation, however, the inflaton potential does not present a minimum, which raises the issue of how to address the problem of reheating. In order to address last problem and as a first approach to further research, in section \ref{wi} we discussed how in the warm inflation scenario the radiation-dominated phase is achieved without introducing the reheating phase for this quasi-exponential Hubble function. We hope to return to this point in the near future.


\begin{acknowledgments}
N.V. was funded by Comisi\'on Nacional
de Ciencias y Tecnolog\'ia of Chile through FONDECYT Grant N$^{\textup{o}}$
3150490.
\end{acknowledgments}


\end{document}